\RequirePackage{fix-cm}
\documentclass[twocolumn]{svjour3}          % twocolumn
\smartqed  % flush right qed marks, e.g. at end of proof
\usepackage{graphicx}
\usepackage{bm}
\journalname{jcamd}

\usepackage[numbers]{natbib}
\usepackage{booktabs}
\usepackage{comment,footnote}
\usepackage{multirow}
\usepackage{multicol}
\usepackage{color}
\usepackage{soul}
\usepackage{siunitx}
\usepackage{hyperref}

\begin{document}

\title{Taming Multiple Binding Poses in Alchemical Binding Free Energy Prediction: the $\beta$-cyclodextrin Host-Guest SAMPL9 Blinded Challenge }
%\title{Application of the Alchemical Transfer and Potential to the $\beta$-cyclodextrin Host-Guest SAMPL9 Blinded Challenge}

\titlerunning{SAMPL9 with ATM}        % if too long for running head

\author{Sheenam Khuttan \and
        Solmaz Azimi    \and
        Joe Z. Wu       \and
        Sebastian Dick$^\dagger$\thanks{$^\dagger$Present affiliation: D.E.\ Shaw Research, New  York, NY}  \and
        Chuanjie Wu$^\ddagger$\thanks{$^\ddagger$Present affiliation: Schr\"{o}dinger Inc, New York, NY}     \and
        Huafeng Xu$^\S$\thanks{$^\S$Present affiliation: Atomapper Inc, New York, NY} \and
        Emilio Gallicchio$^\ast$
}

\institute{
S. Khuttan \at
              Department of Chemistry, Brooklyn College of the City University of New York \\
              PhD Program in Biochemistry, Graduate Center of the City University of New York
           \and
S. Azimi \at
              Department of Chemistry, Brooklyn College of the City University of New York \\
              PhD Program in Biochemistry, Graduate Center of the City University of New York
              \and
J. Z. Wu \at
              Department of Chemistry, Brooklyn College of the City University of New York \\
              PhD Program in Chemistry, Graduate Center of the City University of New York
             \and
S.    Dick \at
            Roivant Sciences, New York, NY
            \and
C. Wu       \at
            Roivant Sciences, New York, NY
            \and
H. Xu   \at
             Roivant Sciences, New York, NY
             \and
E. Gallicchio \at
              Department of Chemistry, Brooklyn College of the City University of New York \\
              PhD Program in Chemistry, Graduate Center of the City University of New York \\
              PhD Program in Biochemistry, Graduate Center of the City University of New York \\
              \email{egallicchio@brooklyn.cuny.edu}
           }

\date{}

\maketitle

\begin{abstract}

We apply the Alchemical Transfer Method (ATM) and a bespoke fixed partial charge force field to the SAMPL9 bCD host-guest binding free energy prediction challenge that comprises a combination of complexes formed between five phenothiazine guests and two cyclodextrin hosts. Multiple chemical forms, competing binding poses, and computational modeling challenges pose significant obstacles to obtaining reliable computational predictions for these systems. The phenothiazine guests exist in solution as racemic mixtures of enantiomers related by nitrogen inversions that bind the hosts in various binding poses, each requiring an individual free energy analysis. Due to the large size of the guests and the conformational reorganization of the hosts, which prevent a direct absolute binding free energy route, binding free energies are obtained by a series of absolute and relative binding alchemical steps for each chemical species in each binding pose. Metadynamics-accelerated conformational sampling was found to be necessary to address the poor convergence of some numerical estimates affected by conformational trapping. Despite these challenges, our blinded predictions quantitatively reproduced the experimental affinities for the $\beta$-cyclodextrin host, less one case of the methylated derivative being an outlier. The work illustrates the challenges of obtaining reliable free energy data in in-silico drug design for even seemingly simple systems and introduces some of the technologies available to tackle them. 


\end{abstract}

\section{Introduction}

Developing in-silico methodologies capable of consistent and reliable binding free energy estimates would be a major breakthrough for drug design and other areas of chemical research.\cite{Jorgensen2004,cournia2020rigorous,griego2021acceleration,xu2022slow} With several advanced simulation software packages now routinely used in industry and academia to model binding affinities of protein-drug complexes,\cite{abel2017advancing,ganguly2022amber} the field has made significant strides toward this goal. Alchemical methods have emerged as the industry standard partly because they can target the changes of binding affinities upon specific chemical modifications of the ligands directly.\cite{cournia2017relative,armacost2020novel,schindler2020large} Theoretical and methodological aspects of free energy are continuously refined and improved.\cite{Mey2020Best,lee2020alchemical,macchiagodena2020virtual,khuttan2021alchemical} However, many challenges remain about the quality of potential energy models\cite{qiu2021Parsleydevelopment} and the correct representation of all of the relevant conformations of the molecular systems.\cite{mobley2012let}

The validation of computational predictions with respect to experimental binding affinities has given the community an understanding of the pitfalls of the models, with indications of ways in which they can be avoided.\cite{wang2015accurate,schindler2020large} Blinded validations, such as the Statistical Assessment of the Modeling of Proteins and Ligands (SAMPL),\cite{geballe2010sampl2} have played a particularly important role in this process.\cite{mobley2014blind,GallicchioSAMPL4,azimi2022application,amezcua2022overview} Because computational predictions are formulated without prior knowledge of experimental results, the evaluation of the models' relative performance is free of implicit biases and reflects more faithfully the expected reliability of the computational models in actual research and discovery settings. 

Many SAMPL challenges include host-guest systems are considered to be straightforward, as well as more approachable, theoretically and experimentally, than macromolecular systems in terms of testing for reliability in free energy prediction tools.\cite{pal2016SAMPL5,yin2017overview} In this work, we report our findings in tackling the SAMPL9 bCD challenge set, which includes the binding of five phenothiazine-based drugs\cite{guerrero2008complexation} to the $\beta$-cyclodextrin host and its methylated derivative.\cite{SAMPL9bCDrepo} Molecular complexes of $\beta$-cyclodextrin (bCD) are well-known and are used in a variety of biomedical and food science applications.\cite{bertrand1989substituent} They are extensively modeled\cite{chen2004calculation,Wickstrom2013,henriksen2017evaluating,he2019role,rizzi2020sampl6,wu2021alchemical} and thus provide a familiar testing ground for computational models. However, as we will show, the binding equilibrium between phenothiazines and cyclodextrin hosts is far from straightforward and requires deploying the most advanced computational tools and methods in our arsenal. As also discussed in later sections, handling conformational heterogeneity in the form of chirality and multiple binding poses has been the greatest challenge in our computational protocol.

This paper is organized as follows. We first describe the above molecular systems in detail to illustrate how these exist in equilibrium as a mixture of many conformations, each with its distinct binding characteristics. We then review the Alchemical Transfer Method (ATM)\cite{wu2021alchemical,azimi2022relative} and the FFEngine bespoke force field parameterization used here to estimate the binding free energies of the cyclodextrin complexes. We describe the extensive alchemical process involving absolute as well as relative binding free energy calculations to obtain the binding constants of each pose in the host-guest systems and how these constants are combined\cite{Gallicchio2011adv} to yield values comparable to the experimental readouts. Reaching convergence for some complexes involving slow intramolecular degrees of freedom required advanced metadynamics-based conformational sampling strategies,\cite{bussi2018metadynamics,eastman2017openmm} which we incorporated into the alchemical binding free energy calculations. This significant intellectual and computational effort resulted in converged binding free energy estimates with a very good experimental agreement for the bCD host. The effort also illustrates the major challenges inherent in modeling complex molecular binding phenomena as well as the theories and technologies available to tackle these challenges. 

\section{\label{sec:molsys}Molecular Systems}

\begin{figure*}
    \centering
    \includegraphics{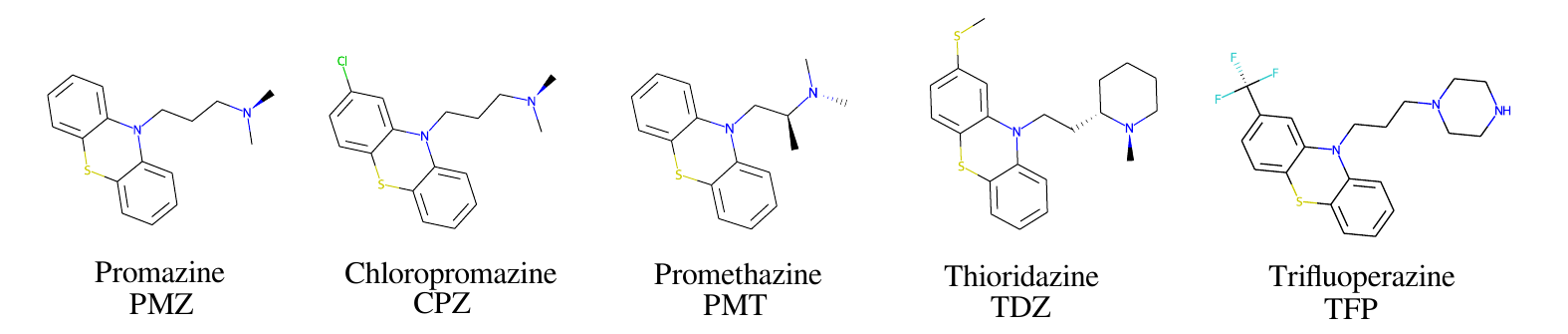}
    \caption{The phenothiazine molecular guests included in the SAMPL9 $\beta$-cyclodextrin challenge.}
    \label{fig:guests}
\end{figure*}

\begin{figure}
    \centering
    \includegraphics[scale=0.7]{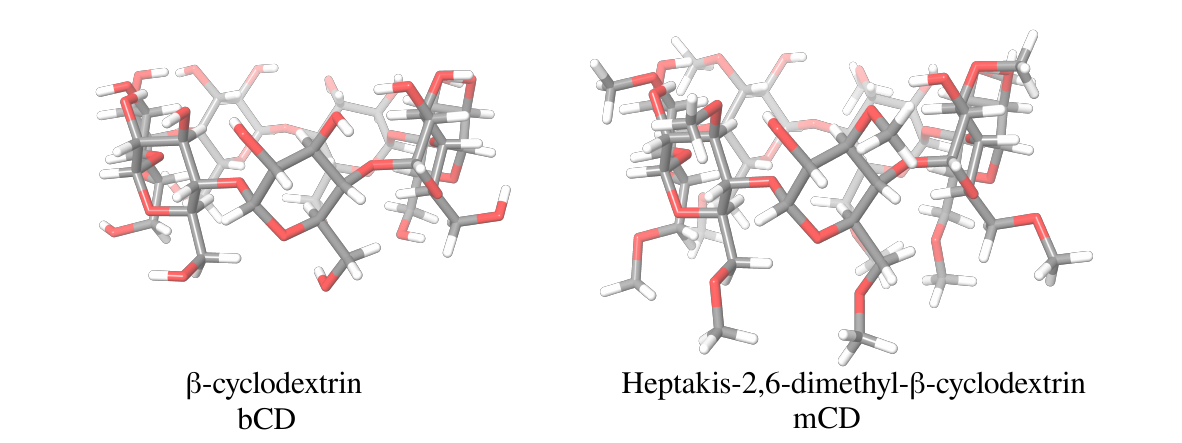}
    \caption{The $\beta$-cyclodextrin (left) and the heptakis-2,6-dimethyl-$\beta$-cyclodextrin (right) molecular hosts included in the SAMPL9 $\beta$-cyclodextrin challenge. The top face $\beta$-cyclodextrin is surrounded by primary hydroxyl groups and the bottom face by secondary hydroxyl groups. The corresponding faces of     heptakis-2,6-dimethyl-$\beta$-cyclodextrin are partially or totally methylated relative to $\beta$-cyclodextrin.}
    \label{fig:hosts}
\end{figure}

The bCD SAMPL9 challenge concerned the binding of five phenothiazine drugs (Figure \ref{fig:guests})\cite{dahl1986phenothiazine} to $\beta$-cyclodextrin (hereafter bCD) and a modified $\beta$-cyclodextrin (hereafter mCD) (Figure \ref{fig:hosts}). The guests\cite{guerrero2008complexation} share a 3-ring phenothiazine scaffold with a variable alkylamine sidechain on the nitrogen atom of the central ring. Unlike the other guests, PMT's alkylamine sidechain is branched at a chiral center. The CPZ, TDZ, and TFP guests also have a variable aromatic substituent on one of the phenyl rings of the phenothiazine scaffold

The $\beta$-cyclodextrin host (Figure \ref{fig:hosts}) is a cyclic oligosaccharide of seven D-glucose monomers forming a binding cavity with a wide opening surrounded by secondary hydroxyl groups (top in Figure \ref{fig:hosts}) and a narrower opening (bottom) surrounded by primary hydroxyl groups. Hence, we will refer to the wide opening as the secondary face of bCD and the narrow opening as the primary face. The second host, heptakis-2,6-dimethyl-$\beta$-cyclodextrin (mCD, Figure \ref{fig:hosts}), is a derivative of $\beta$-cyclodextrin in which all of the primary hydroxyl groups and half of the secondary ones are methylated, affecting the size, accessibility, and hydrophobicity of the binding cavity. Although mCD does not have secondary and primary hydroxyl groups, for simplicity, we will refer to the two openings of mCD as secondary and primary faces by analogy with bCD. Being composed of chiral monomers, bCD and mCD are themselves chiral molecules with potentially different affinities for the enantiomers of optically active guests.\cite{rekharsky2000chiral}

\subsection{Multiple Chemical Species of the Guests}

\begin{figure}
    \centering
    \includegraphics[scale=0.6]{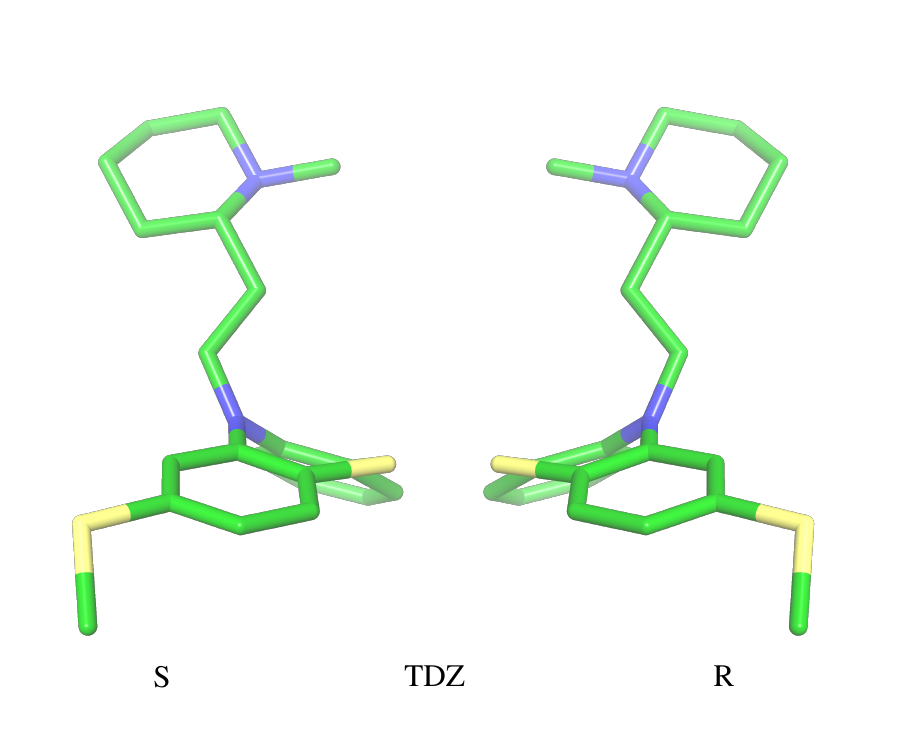}
    \caption{Illustration of the two conformational enantiomers of the TDZ guest. Similarly to the CPZ and TFP guests, chirality is induced by the aromatic substituent (a thioether here). The unsubstituted guests PMZ and PMT do not possess conformational chirality.}
    \label{fig:pheno-chirality}
\end{figure}

 The amine group of the alkylamine sidechain is expected to be largely protonated in solution and the host-guest complex at the pH of the experiment. However, the two tautomers of the protonated piperazine group of the TFP guest are likely exist at appreciable concentrations and can contribute to host binding to different extents. Similarly, in the case of TDZ, protonation of the alkyl nitrogen produces two enantiomers that can interact differently with the cyclodextrin hosts. Moreover, rather than being planar, the phenothiazine ring system is bent at the connecting central ring  with conformations with both positive and negative bends present in equal amounts in solution (Figure \ref{fig:pheno-chirality}). As illustrated for TDZ in Figure \ref{fig:pheno-chirality}, when an aromatic substituent is present, the species with positive and negative bend are conformational enantiomers, each with the potential to interact differently with the cyclodextrin hosts.\cite{rekharsky2000chiral,quinton2018confining}
 
 The experimental binding assay reports an average over the contributions of the various chemical species of the guests. However, because interconversions between species cannot occur in molecular mechanics simulations or occur too slowly relative to the molecular dynamics timescales, to obtain a binding affinity estimate comparable with the experimental observations, it is necessary to model the binding of each relevant species individually and combine the results.\cite{Jayachandran2006} In this work, we have considered the two conformational enantiomers for each guest (including those of PMZ and PMT with the unsubstituted phenothiazine scaffold compounds to test for convergence), plus the two chiral forms of the protonated alkyl nitrogen of TDZ and the two forms of TFP protonated at the distal and proximal alkyl nitrogens, for a total of 14 guest species. We labeled the seven species with R chirality of the phenothiazine scaffold as PMZ1R, CPZ1R, PMT1R, TDZNR1R, TDZNS1S, TFP1aR, and TFP1bR, where the first part of the label identifies the compound, followed by the net charge ($+1$ for all the species considered) with `a' and `b' label identifying the distal and proximal protonated forms of TFP respectively, plus `NR' and `NS' labels to distinguish the R and S chiral forms of the protonated alkyl nitrogen of TDZ. The last letter identifies the chirality of the phenothiazine scaffold so that the seven species with S chirality are named PMZ1S, CPZ1S, etc.

\subsection{Multiple Binding Poses}

\begin{figure}
    \centering
    \includegraphics[scale=0.5]{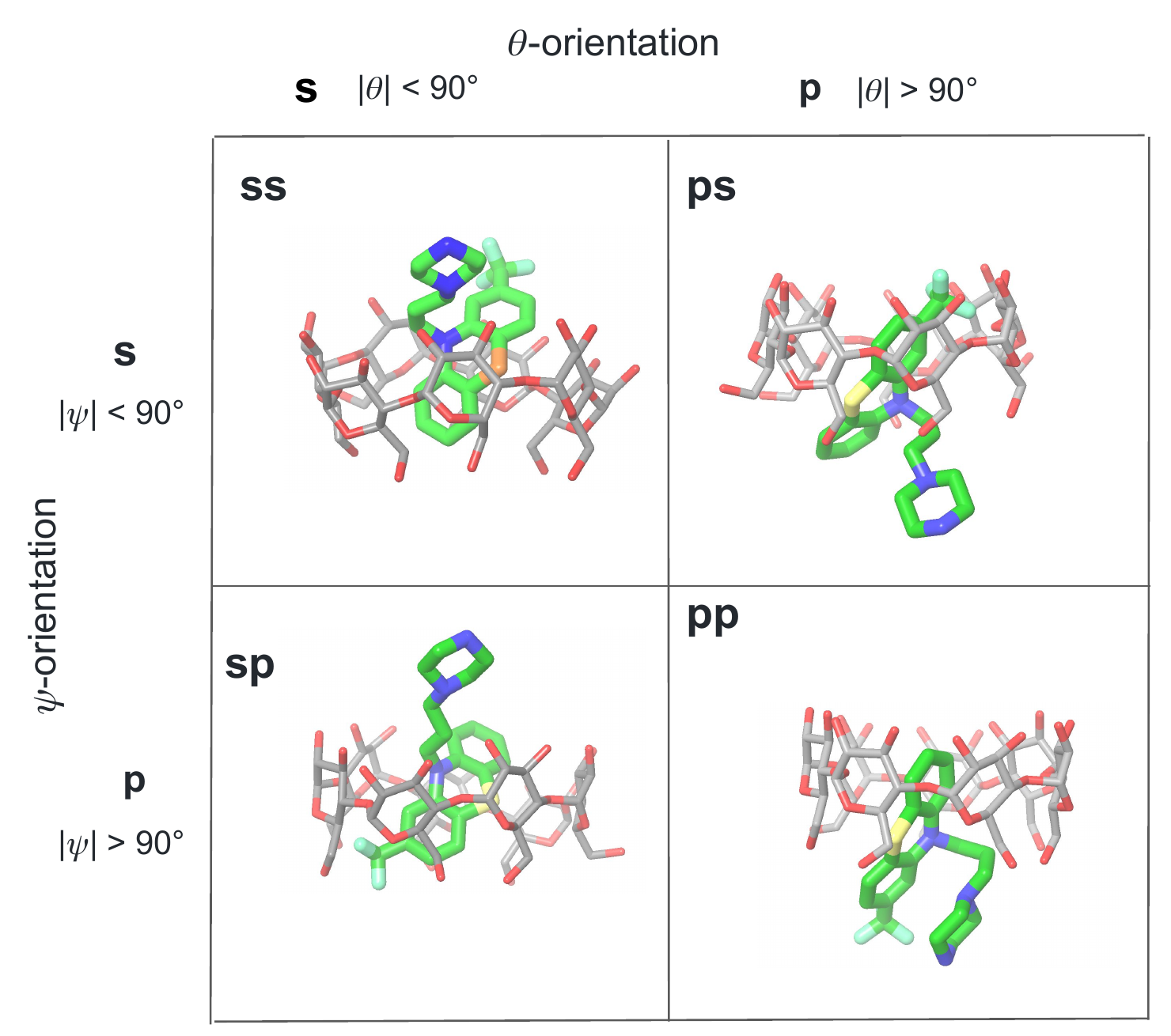}
    \caption{Illustration of the classification of the four binding poses of the phenothiazine/cyclodextrin complexes based on the polar and twist angles introduced in Figure \ref{fig:posedef}. Poses are labeled as `ss', `sp', etc.\ where `s' refers to the secondary face of the host and `p' to the primary face of the host. The first letter of the label refers to the orientation of the alkylamine sidechain that can protrude out from the secondary face (poses `ss' and `sp') or from the primary face of the host (poses `ps' and `pp'). Similarly, the second letter refers to the position of the aromatic substituent protruding out from either the primary or secondary faces of the host.  }
    \label{fig:poses}
\end{figure}

The guests included in the SAMPL9 bind the cyclodextrin hosts in four distinct binding modes (Figure \ref{fig:poses}). To identify the binding modes, we will refer to the narrow opening of the $\beta$-cyclodextrin circled by primary hydroxyl groups as the primary face of the host (the bottom opening in Figure \ref{fig:hosts}). Similarly, the wider opening (top in Figure \ref{fig:hosts}) surrounded by secondary hydroxyl groups will be referred to as the  secondary face of the hosts. The guests can bind the cyclodextrin hosts with the alkylamine sidechain pointing towards the host's secondary (denoted by `s') or primary (denoted by `p') faces (Figure \ref{fig:poses}). Each of these poses  is further classified in terms of the position of the aromatic substituent, which can be at either the secondary or primary faces of the host. Hence the binding modes of the guest/cyclodextrin complexes are labeled: `ss', `sp', `ps', and `pp', where the first letter refers to the position of the alkylamine sidechain and the second to the position of the aromatic substituent (Figure \ref{fig:poses}). 

The binding mode labels are combined with the labels discussed above that identify the chemical form of the guest to obtain the labels for each form of the guest in each binding mode. For example, the guest PMT with $+1$ charge with R chirality in the `ss' binding mode is labeled as PMT1Rss.  

For the purpose of the alchemical calculations, the binding modes are defined geometrically in terms of the polar angle $\theta$ and the azimuthal angle $\psi$ illustrated in Figure \ref{fig:posedef}. $\theta$ is the angle between the molecular axes of the host and the guest and determines the orientation of the alkylamine sidechain relative to the host. The molecular axis of the cyclodextrin host (labeled $z$ in Figure \ref{fig:posedef}) is oriented from the primary to the secondary faces going through the centroid of the atoms lining the faces (see Computational Details). The molecular axis of the guests goes from the sulfur and nitrogen atoms of the central phenothiazine ring. The angle $\psi$ describes the rotation around the molecular axis of the guest and determines the position of the aromatic substituent. The `sp' binding mode, for example, is defined by $0 \le \theta \le 90^\circ$ and $90^\circ \le \psi \le 180^\circ$ (see Figure \ref{fig:poses} and Computational Details).

\section{\label{sec:methods}Theory and Methods}

\subsection{Design of the Alchemical Process }

\begin{figure}
    \centering
    \includegraphics[scale=0.9]{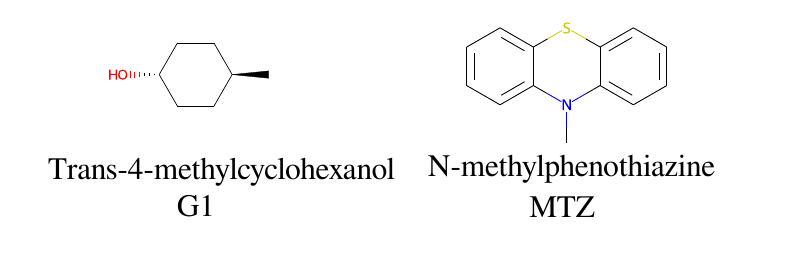}
    \caption{The structures and abbreviations of the molecular guests used as intermediate compounds in the alchemical process. }
    \label{fig:otherguests}
\end{figure}

\begin{figure}
    \centering
    \includegraphics[scale=0.7]{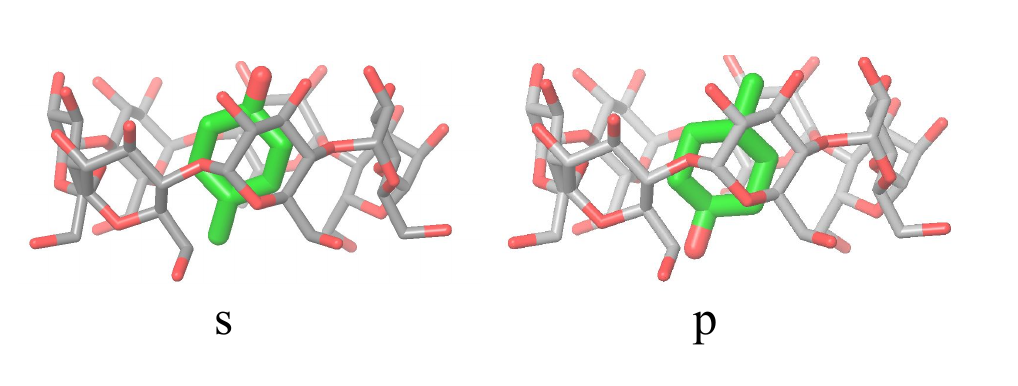}
    \caption{The `s' and `p' binding modes of the G1/$\beta$-cyclodextrin complex. The `s' mode, in which the hydroxyl group points towards the secondary face of the host, is used as a starting species for the `ss' and `sp' binding modes of the phenothiazine/cyclodextrin complexes. The `p' mode, which points towards the primary face, is the starting species for the `ps' and `pp' modes (Figure \ref{fig:poses}).}
    \label{fig:g1}
\end{figure}

The alchemical calculations aim to estimate the guests' absolute binding free energies (ABFEs) to each host. Direct alchemical ABFE calculations failed to reach convergence for these systems partly due to the relatively large sizes of the guests and partly because of the slow conformational reorganization of the cyclodextrin hosts from a closed apo state to an open guest-bound state.\cite{wickstrom2016parameterization,he2019role} To overcome these obstacles, we adopted a step-wise process made of a series of relative binding free energy calculations (RBFE) starting from the ABFE of a small guest that could be reliably estimated. Specifically, we obtained the ABFE of trans-4-methylcyclohexanol (Figure \ref{fig:otherguests})--the G1 guest of the SAMPL7 bCD challenge\cite{amezcua2021sampl7}--for the secondary and primary binding modes to each host. We defined the `G1s' binding mode of the G1 guest as the one in which the hydroxyl group points toward the secondary face of the cyclodextrin host, while it points to the opposite face in the `G1p' mode (Figure \ref{fig:g1}).

Each binding mode of the complex with G1 was then alchemically converted to an intermediary phenothiazine guest (N-methylphenothiazine, or MTZ, in Figure \ref{fig:otherguests}), similar to the SAMPL9 phenothiazine derivatives with a small methyl group replacing the large alkylamine sidechains. Even though MTZ does not have conformational chirality (Figure \ref{fig:pheno-chirality}), we treated its S and R enantiomers individually to test the convergence of the RBFE estimates for each binding pose. We used atom indexes to distinguish the S and R enantiomers of these symmetric guests. Calculations were conducted to obtain the RBFEs from the G1s to the MTZRsp, MTZRss, MTZSsp, and MTZSss binding poses of the complexes of MTZ with bCD and mCD, and from G1p to the MTZRps, MTZRpp, MTZSps, and MTZSpp of the same complexes, all independently and from different starting conformations. The MTZRsp, MTZRss, MTZSsp, and MTZSss complexes are equivalent and should yield the same RBFE values within uncertainty. Similarly, the MTZRps, MTZRpp, MTZSps, and MTZSpp should yield equivalent RBFEs but distinguishable from those of the MTZRsp, MTZRss, MTZSsp, MTZSss group.

\begin{figure*}
    \centering
    \includegraphics[scale=0.85]{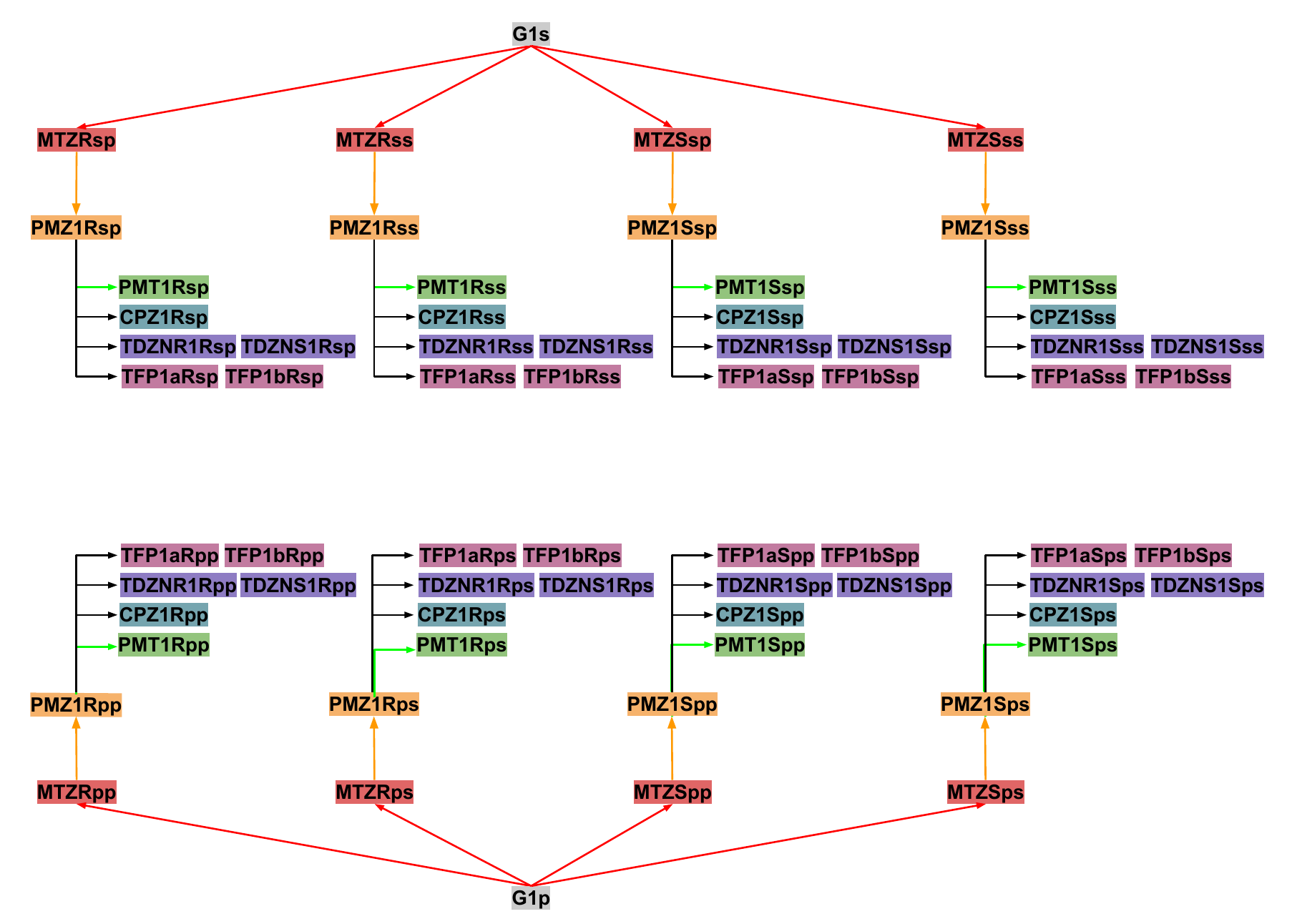}
    % https://docs.google.com/drawings/d/1OfxGQJUSvByHuECZXHAEHMdzv2TAgUqyOviEmqeVcbQ/edit
    \caption{The map of relative binding free energy calculations to obtain the binding free energies of each pose of each guest starting from the absolute binding free energy of the corresponding poses of the G1 guest. Nodes of the same color contribute to the binding free energy estimate of one of the five guests: PMZ (yellow), PMT (green), CPZ (cyan), TDZ (violet), and TFP (purple).}
    \label{fig:alchemical-process}
\end{figure*}

Next, RBFEs were obtained for each complex of MTZ to the corresponding complex of PMZ. For example, the MTZRsp binding pose of the MTZ complex with bCD and mCD were converted to the PMZ1Rsp binding pose of the corresponding complexes between PMZ and the hosts. Finally, the RBFEs between each pose of PMZ and the corresponding binding poses of the other guests were obtained. During this process, we monitored convergence by looking at the discrepancy between the RBFEs corresponding to the equivalent symmetric poses of the achiral PMZ and PMT guests. The overall alchemical process to obtain the ABFEs of the SAMPL9 phenothiazine guests is summarized in Figure \ref{fig:alchemical-process}.

\begin{figure}
    \centering
    \includegraphics[scale=0.4]{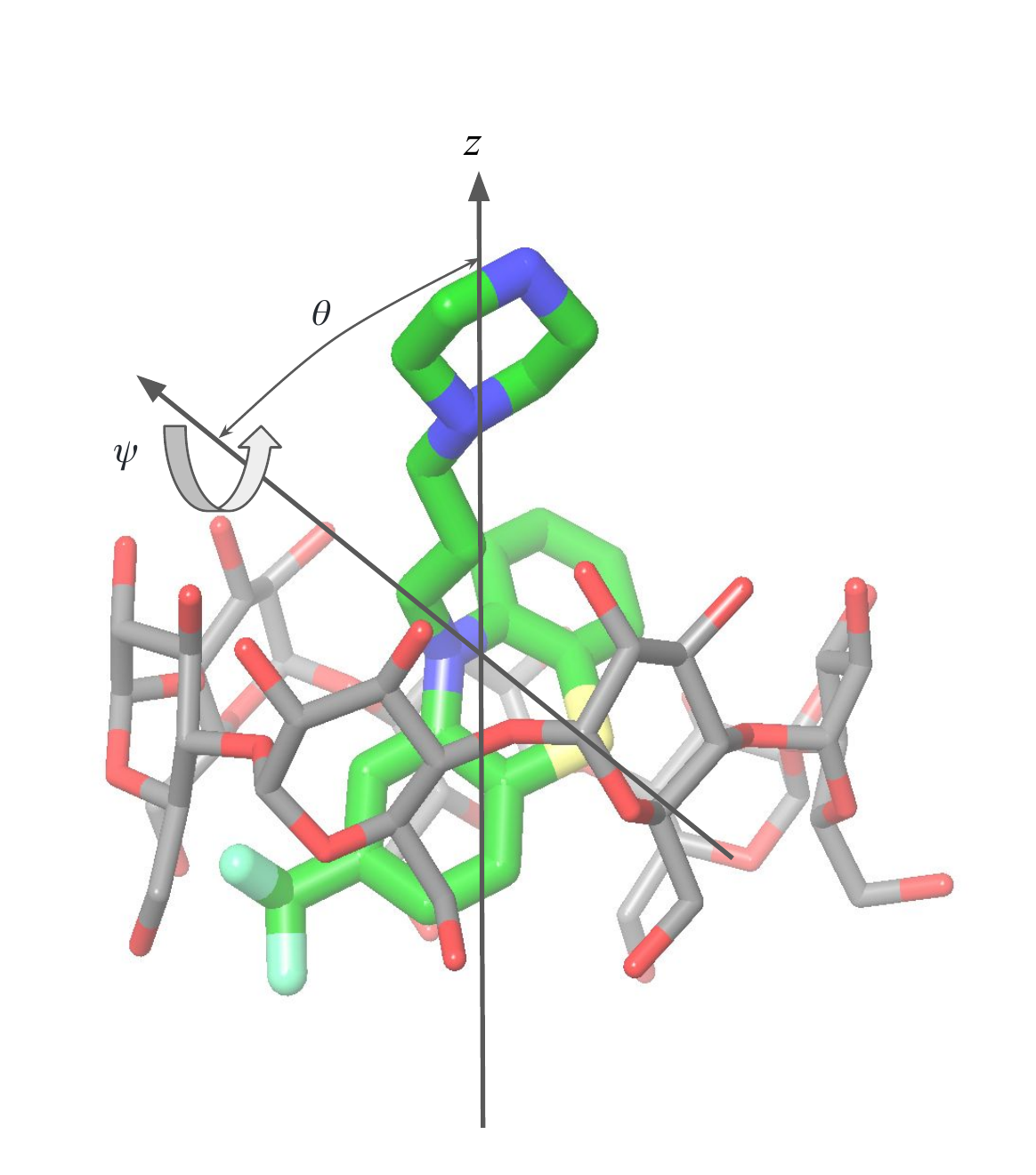}
    \caption{Illustration of the geometrical definition of the binding poses of the phenothiazine/cyclodextrin complexes. The definition is based on the polar ($\theta$) and twist ($\psi$) angles of the molecular axis of the guest with respect to the coordinate frame of the host, which includes the $z$-axis that runs from the primary to the secondary face of the host. See Computational Details for the specific definition of the guests' and hosts' coordinate frames.}
    \label{fig:posedef}
\end{figure}

\subsubsection{Free Energy of Binding for Complexes with Multiple Binding Modes }
 
The observed binding constant $K_b$ of the complex $RL$ of a receptor $R$ with a ligand $L$ present in forms or poses $L_i$, $i = 1, 2, \ldots$ is the weighted average of the binding constant $K_b(i)$ for each form with weights equal to the relative population $P_0(i)$ of each form in solution\cite{Jayachandran2006,Gallicchio2011adv}
 \begin{equation}
     K_b = \sum_i P_0(i) K_b(i)
     \label{eq:combination-formula}
 \end{equation}
Statistical mechanics-based derivations of this formula, which we refer to as the Free Energy Combination Formula, are available.\cite{Jayachandran2006,Gallicchio2011adv} The Free Energy Combination Formula can also be derived using elementary notions as follows: $K_b = C^\circ \frac{[RL]}{[R][L]} = \sum_i C^\circ \frac{[RL_i]}{[R][L]} =  \sum_i \frac{[L_i]}{[L_i]} C^\circ \frac{[RL_i]}{[R][L]}$ and $C^\circ \frac{[RL_i]}{[R][L]} = K_b(i)$ and  $\frac{[L_i]}{[L]} = P_0(i)$, where $C^\circ = 1$ mol/L, $[RL] = \sum_i [RL_i]$ is the total molar concentration of the complex and $[RL_i]$ is the concentration of mode $i$ of the complex. Similar definitions apply to the concentrations of the ligand $[L]$ and $[L_i]$, and $P_0(i) = [L_i]/[L]$ is the population of mode $i$ of the ligand in solution.  

Moreover, as also shown in the Appendix, the fractional contribution of binding mode $i$ to the overall binding constant is the population, $P_1(i)$ of mode $i$ of the complex:\cite{Gallicchio2011adv}
 \begin{equation}
     P_1(i) = \frac{ P_0(i) K_b(i)}{K_b}
     \label{eq:p1}
 \end{equation}
 Below we used this property to infer the probability of occurrence of each mode of the host-guest complexes.   
 
In this specific application, the binding modes refer to the `ss', `sp', etc. orientations of the R and S enantiomers of each guest. We individually obtained the binding mode-specific binding constants $K_b(i)$ for each binding mode. In the corresponding alchemical simulations, the orientation of the ligand in the binding site is set by restraining potentials based on the $\theta$ and $\psi$ angles (see Figure \ref{fig:posedef} and Computational Details). These orientations are equally likely in solutions. We also assume an equal likelihood of the R and S conformational enantiomers of the guests in solution, leading to $P_0(i) = 1/8$ for each pose of each guest.  The TDZ and TFP guests have twice as many poses due to point chirality and multiple protonation states of their alkylamine sidechain that we also assume to be equally likely in solution. Hence, we set $P_0(i) = 1/16$ for each state of the TDZ and TFP guests. 

\subsection{The Alchemical Transfer Method}

The Alchemical Transfer Method (ATM) is based on a coordinate displacement perturbation of the ligand between the receptor-binding site and the explicit solvent bulk and a thermodynamic cycle connected by a symmetric intermediate in which the ligand interacts with the receptor and solvent environments with equal strength.\cite{wu2021alchemical,azimi2022application} The perturbation energy $u$ for transferring the ligand from the solution to the binding site is incorporated into a $\lambda$-dependent alchemical potential energy function of the form
\begin{equation}
U_\lambda(x) = U_0(x) + W_\lambda(u)
\label{eq:alch-func}
\end{equation}
where $x$ represents the system's coordinates, $U_0(x)$ is the potential energy function that describes the unbound state, and $W_\lambda$ is the softplus alchemical potential\cite{khuttan2021alchemical,wu2021alchemical,azimi2022application} such that the system's potential energy function is transformed from that of the unbound state to that of the bound state as $\lambda$ goes from $0$ to $1$. This alchemical process computes the excess component of the free energy of binding. The ideal term $-k_B T \ln C^\circ V_{\rm site}$, where $C^\circ = 1$ mol/L and $V_{\rm site}$ is the volume of the binding site is added in post-processing to obtain the standard free energy of binding.

In this work, we used the strategy above to compute the absolute binding free energy (ABFE) of the G1 guest in two different poses. The binding free energies of the phenothiazine guests are obtained by a series of relative binding free energy (RBFE) calculations starting from G1 (Figure \ref{fig:alchemical-process}). The alchemical RBFE implementation of the ATM method\cite{azimi2022relative} is similar to ABFE implementation except that one ligand of the pair under investigation is transferred from the solution to the binding site while the other ligand is transferred in the opposite direction. The receptor and the two ligands are placed in a single solvated simulation box in such a way that one ligand is bound to the receptor and the other is placed in an arbitrary position in the solvent bulk. Molecular dynamics simulations are then conducted with a $\lambda$-dependent alchemical potential energy function that connects, in an alchemical sense, the state of the system where one guest is bound to the receptor and the other in solution, to the state where the positions of the two guests are reversed. The free energy change of this process yields the relative binding free energy of the two guests. To enhance convergence, the two ligands are kept in approximate alignment to prevent the one in solution to reorient away from the orientation of the bound pose. We have shown mathematically that the alignment restraints implemented in ATM do not bias the binding free energy estimates.\cite{azimi2022relative}

In this work, we employed metadynamics conformational sampling to obtain converged RBFE estimates for the PMT guest. Well-tempered metadynamics\cite{barducci2008well} is a well-established enhanced sampling technique to sample rare events during MD simulations when they are separated from other metastable states by large energy barriers. The technique uses a bias potential, $U_{\rm bias}$,  that lowers energy barriers along a slow degree of freedom. In this work, the metadynamics biasing potential is obtained along a dihedral angle, $\varphi$, of the alkylamine sidechain of PMT (see Computational Details) from a simulation in a water solution, using OpenMM's well-tempered metadynamics implementation by Peter Eastman.\cite{eastman2017openmm} The alchemical binding free energy calculation is then conducted with the biasing potential, $U_{\rm bias}(\varphi)$, added to the alchemical potential energy function in Eq.\ (\ref{eq:alch-func}). The resulting binding free energy estimate is then unbiased using a book-ending approach\cite{hudson2019use} by computing the free energy differences of the system without the biasing potential from samples collected with the biasing potential at the endpoints of the alchemical path. In this work, we used a simple unidirectional exponential averaging formula
\[
-k_B T \ln \langle \exp(U_{\rm bias}/k_B T) \rangle_{\rm bias}
\]
to evaluate the free energy corrections for unbiasing. Due to the larger excursions of the dihedral angle with metadynamics, the unbiased ensemble is a subset of the biased ensemble and the exponential averaging estimator converges quickly in this case.

\subsection{Force Field Parametrization}

Force field parameters were assigned to the hosts and the guests using an in-house development FFEngine workflow at Roivant Discovery. FFEngine is a workflow for the bespoke parametrization of ligands with the Amber force field functional form.\cite{wang2006automatic} The partial charges were derived from GFN2-xTB/BCC with pre-charges from semi-empirical QM method GFN2-xTB,\cite{bannwarth2019gfn2} and bond charge correction (BCC) parameters fitted to the HF/6-31G* electrostatic potential (ESP) with the COSMO implicit solvation model from a 50,000 drug-like compounds dataset. The  ESP with an implicit solvation model was deemed necessary for these highly polar and charged host-guest systems even though it is expected to yield a fixed charge model slightly more polarized than the default GAFF2/Amber one.

\subsection{Computational Details }

The guests were manually docked to the hosts in each binding pose using Maestro (Schr\"{o}dinger, Inc.) in each of the four binding poses. A flat-bottom harmonic positional restraint with a force constant $k_c = 25$ kcal/mol/\AA$^2$ and a tolerance of $4$ \AA\ was applied to define the binding site region.\cite{Gilson:Given:Bush:McCammon:97,Gallicchio2011adv} For this purpose, the centroid of the guest was taken as the center of the central ring of the phenothiazine core, and the centroid of the cyclodextrin host was defined as the center of the oxygen atoms forming the ether linkages between the sugar monomers. Boresch-style\cite{Boresch:Karplus:2003} orientational restraints were imposed to keep each complex in the chosen binding pose. These were implemented as flat-bottom restraints acting on the $\theta$ and $\phi$ angles in Figure \ref{fig:posedef} with a force constant of $k_a = 0.05$ kcal/mol/deg$^2$, and centers and tolerances tailored for each pose. For example, the orientational restraints for the `sp' pose are centered on $\theta = 0^\circ$ and $\phi = 180^\circ$ with $\pm 90^\circ$ tolerances for both.  The $\theta$ angle is defined as the angle between the $z$-axis of the host, defined as the axis going through the centroid of the oxygen atoms of the primary hydroxyl groups and the centroid of the oxygen atoms of the secondary hydroxyl groups, and the molecular axis of the guest, defined as the axis going through the S and N atoms of the central ring of the phenothiazine core. The $\phi$ angle is defined as the dihedral angle between the plane formed by the C1-N-S triplet of atoms of the phenothiazine core of the guest and the plane spanned by the $z$-axis of the host and the molecular axis of the guest, where C1 represents the carbon atom of the phenothiazine host with the aromatic substituent. Very loose flat-bottom harmonic positional restraints ($4$ \AA\ tolerance and $25$ kcal/mol/\AA$^2$ force constant) were applied to the ether linkages oxygen atoms of the cyclodextrins to keep the hosts from wandering freely in the simulation box.  The ATM displacement vector was set to $(-30,0,0)$ \AA\ .

Force field parameters were assigned as described above. In RBFE calculations, the second ligand in the pair was placed in the solvent by translating it along the displacement vector. The resulting system was then solvated using tleap\cite{AmberTools} in a TIP3P box with a 10 \AA\ solvent buffer and sodium and chloride counterions to balance the negative and positive charges, respectively. The systems were energy minimized, thermalized, and relaxed in the NPT ensemble at 300 K and 1 atm pressure. Annealing of the systems to $\lambda = 1/2$ in the nVT ensemble followed to obtain initial structures to initiate the parallel replica exchange ATM calculations. Alchemical calculations were conducted with the OpenMM 7.7 MD engine on the CUDA platform with a 2 fs time-step, using the AToM-OpenMM software.\cite{AToM-OpenMM} Asynchronous Hamiltonian Replica Exchange\cite{gallicchio2015asynchronous} in $\lambda$ space was attempted every 20 ps and binding energy samples were collected at the same frequency.   The ATM-RBFE employed 22 $\lambda$ states distributed between $\lambda$ = 0 to $\lambda$ = 1 using the non-linear alchemical soft-plus potential and the soft-core perturbation energy with parameters $u_{\rm max} = 200$ kcal/mol, $u_c = 100$ kcal/mol, and $a=1/16$.\cite{khuttan2021alchemical} The input files of the simulations are provided in our lab's GitHub repository (https://github.com\-/GallicchioLab\-/SAMPL9-bCD). We collected approximately 20 ns trajectories for each replica corresponding to approximately 440 ns for each of the 64 alchemical steps for each host (Figure \ref{fig:alchemical-process}). Overall, we simulated the systems for over $6$ $\mu$. UWHAM\cite{Tan2012} was used to compute binding free energies and the corresponding uncertainties after discarding the first half of the trajectory. 

To obtain the torsional flattening biasing potential, we simulated the PMT guest in solution with metadynamics over the (C-N-C-C) alkylamine sidechain torsional angle highlighted in Figure \ref{fig:pmf-torsion}. A well-tempered metadynamics bias factor of 8 was used, with a Gaussian height of 0.3 kcal/mol and width of $10^\circ$.\cite{barducci2008well}  The bias potential was collected for 20 ns, updating it every 100 ps. The resulting potential of mean force is shown in Figure \ref{fig:pmf-torsion}. The metadynamics-derived biasing potential was used in all the RBFE calculations involving the PMT guest to accelerate the sampling of the slow torsional degree of freedom in question.

\section{Results}

\subsection{Binding Free Energy Predictions }

The calculated binding free energies of the cyclodextrin/phenothiazine complexes obtained by combining the pose-specific binding free energies are listed in Table \ref{tab:dgvsexpt} compared to the experimental measurements. We provide the results of each individual free calculation in the Supplementary Information. The second column of Table \ref{tab:dgvsexpt} reports the  blinded computational predictions submitted to the SAMPL9 organizers and the results of revised predictions (third column) obtained subsequently to correct setup errors and resolve unconverged calculations. Specifically, we uncovered cases where binding poses were misidentified and where centers of ligands and the hosts had reversed chirality during energy minimizations due to close initial atomic overlaps. As discussed below, in the initial predictions, we were also unable to obtain consistent binding free energy predictions for symmetric poses. 

The blinded and revised predictions for the bCD complexes agree with the experiments. The revised predictions, in particular, are all within 1.5 kcal/mol of the experimental measurements and within 1 kcal/mol for four of the five bCD complexes. Although the range of the binding affinities is small, some trends are reproduced and the weakest binder (PMT) is correctly identified. The quality of the predictions for the mCD host is not as good, and it worsened upon revision. The experiments show that the phenothiazine guests bind slightly more strongly to mCD than bCD. However, except for CPZ, the calculations predict significantly weaker binding to mCD relative to bCD. The computed free energies of the mCD complexes are on overage over 2 kcal/mol less favorable than the experimental ones. The revised prediction for the mCD-TDZ complex is particularly poor and fails to identify this complex as the most stable in the set. While a detailed investigation of the sources of the poor prediction for mCD has not been carried out, our model could not have identified the best possible binding poses for this more flexible host. mCD is also more hydrophobic and the energy model may overpredict the reorganization free energy to go from the apo to the holo conformational ensemble for this host.  

\begin{table*}
\caption{\label{tab:dgvsexpt}The binding free energy predictions submitted to the SAMPL9 challenge compared to the revised predictions and the experimental measurements.}
\begin{center}
\sisetup{separate-uncertainty}
\begin{tabular}{l S[table-format = 3.2(2)] S[table-format = 3.2(2)] S[table-format = 3.2]}
 Name & \multicolumn{1}{c}{$\Delta G_b$(SAMPL9)$^{a,b,c}$} & \multicolumn{1}{c}{$\Delta G_b$(ATM)$^{a,b,d}$} & \multicolumn{1}{c}{$\Delta G_b$(expt)$^{a,e}$} \\ [0.5ex] 
\hline
bCD-TDZ & -4.28(90) & -4.56(47) & -5.73 \\ 
bCD-TFP & \multicolumn{1}{S[table-format = 3.2(3)]}{-6.51(111)} & -5.42(99) & -5.09 \\ 
bCD-PMZ & -3.73(48) & -4.03(45) & -5.00 \\ 
bCD-PMT & -2.53(70) & -3.00(47) & -4.50 \\ 
bCD-CPZ & -7.28(92) & -4.64(70) & -5.45 \\ 
 \hline
mCD-TDZ & \multicolumn{1}{S[table-format = 3.2(3)]}{-5.16(140)} & -2.96(68) & -6.50 \\ 
mCD-TFP & -4.14(62) & -3.98(70) & -5.57 \\ 
mCD-PMZ & -2.37(54) & -2.34(55) & -5.08 \\ 
mCD-PMT & -1.80(99) & -1.58(60) & -5.39 \\ 
mCD-CPZ & -5.22(90) & -5.13(88) & -5.43 \\ 
 \hline
\end{tabular}
\begin{flushleft}\small
$^a$In kcal/mol. $^b$Errors are reported as twice the standard deviation. $^c$Blinded computational predictions submitted to the SAMPL9 challenge organizers. $^d$Revised ATM computational predictions. $^e$Provided by the SAMPL9 organizers: https://www.samplchallenges.org.
\end{flushleft}
\end{center}
\end{table*}

\subsection{Binding Mode Analysis}

\begin{figure*}
    \centering
    \includegraphics[scale=0.50]{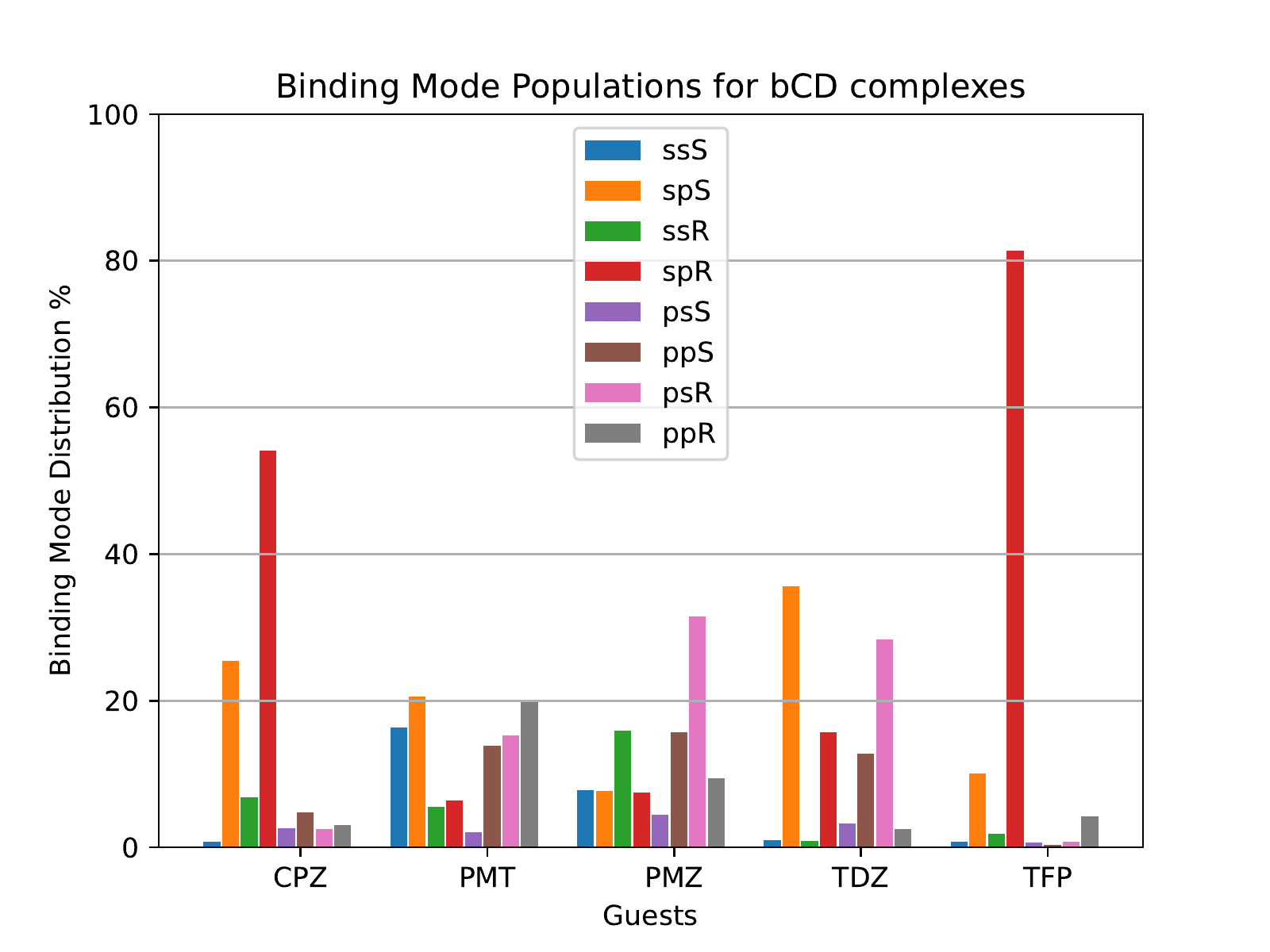} \includegraphics[scale=0.50]{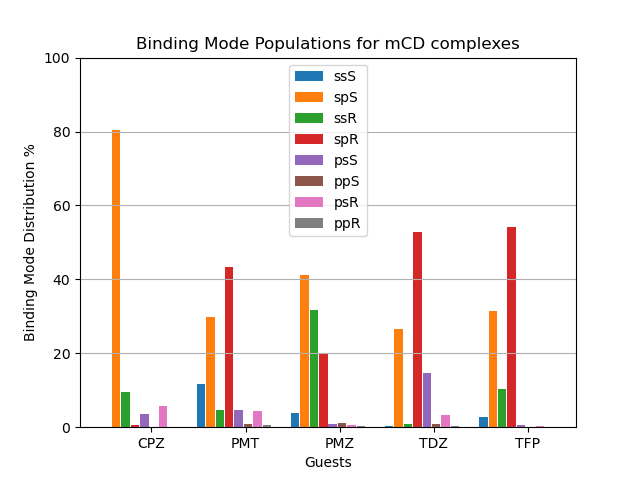}
    \caption{\label{fig:bmode-populations} Binding mode populations of the complexes with bCD (left) and mCD (right).}   
\end{figure*}

We used the binding mode-specific binding constants  we obtained (see Supplementary Information) to infer the population of each binding mode for each complex shown in Figure \ref{fig:bmode-populations}. The results indicate that the complexes visit many poses with appreciable probability. The only exceptions are TFP binding to bCD and CPZ binding to mCD which are predicted to exist with over 75\% probability in only one pose (`sp' in the R configuration in the case of the TFP-bCD complex and `sp' in the S configuration in the case of the CPZ-mCD complex). In general, the guests bind the hosts preferentially in the `sp' and `ss' modes with the alkylamine sidechain placed near the primary face of the hosts (Figure \ref{fig:poses}). This trend is less pronounced for the complexes between PMT, PMZ, and TDZ with bCD, which occur in the `sp'/`ss' and `pp'/`ps' modes with similar frequency, and it is more pronounced for all complexes with mCD which strongly prefer the alkylamine sidechain towards the primary face. Unlike the alkylamine sidechain, the aromatic substituent of the CPZ, TDZ, and TFP guests is preferentially placed towards the secondary face of the cyclodextrin hosts. This is evidenced by the higher probability of the `sp' binding modes (red and green bars in Figure \ref{fig:bmode-populations}) over the `ss' binding modes (blue and yellow).  

Reassuringly, the calculations predict that the populations of the symmetric binding modes of the complexes with the PMZ and PMT guests are more evenly distributed than for the other complexes. Lacking an aromatic substituent (Figure \ref{fig:guests}), the PMZ and PMT guests do not display conformational chirality (Figure \ref{fig:pheno-chirality}). Hence, their `ssS', `spS', `ssR`, and `spR' binding modes are chemically equivalent and should have the same population. Similarly, the binding modes `psS', `ppS', `psR`, and `ppR' of these guests are mutually equivalent. Still, they are distinguishable from the `ssS', `spS', `ssR`, `spR' group by the position of the alkylamine sidechain (Figure \ref{fig:poses}). We used these equivalences to assess the level of convergence of the binding free energy estimates. Although redundant for symmetric poses, we simulated each binding mode of these guests individually, starting from different initial configurations, and checked how close the pose-specific binding free energies varied within each symmetric group. For example, the computed populations of the `ssS', `spS', `ssR`, and `spR' poses of the PMZ-bCD complex vary in a narrow range between $7.5$ and $15.9$\%, indicating good convergence. However, the corresponding populations for the complex with mCD are not as consistent--the 'ssS' mode predicted to be significantly less populated (4\%) than the other modes (20-40\%)--reflecting poorer convergence.

The pose-specific binding free energy estimates probe the chiral binding specificity of the hosts. Except for the TFP guest that is predicted to bind predominantly in the R chiral form (88\% population), bCD shows little chiral preference.  mCD displays a slightly  stronger chiral specificity, with CPZ predicted to bind predominantly in the S form and TFP in the R form.

\subsection{Enhanced Conformational Sampling of the PMT Guest }

As discussed above, the `ssS', `spS', `ssR', and `spR' binding poses of the PMT guest, which lacks an aromatic substituent, are chemically indistinguishable and should yield equivalent pose-specific ABFE estimates. Similarly, the `psS', `ppS', `psR', and `ppR' should yield the same binding free energy within statistical uncertainty. Yet, in our first attempt submitted to SAMPL, our predictions did not achieve the expected consistency (Table \ref{tab:dgpmt1}, second column). In Table \ref{tab:dgpmt1} we show the binding free energy estimates for each PMT binding pose relative to the same pose of PMZ, whose poses are equivalent in the same way as for PMT. For instance, while the four top poses for bCD are expected to yield the same RBFEs, the actual estimates show a scatter of more than $4$ kcal/mol. The other groups of equivalent binding poses of bCD and mCD also show significant scatter, which is indicative of poor convergence. 

\begin{figure}
    \centering
    \includegraphics[scale=0.25]{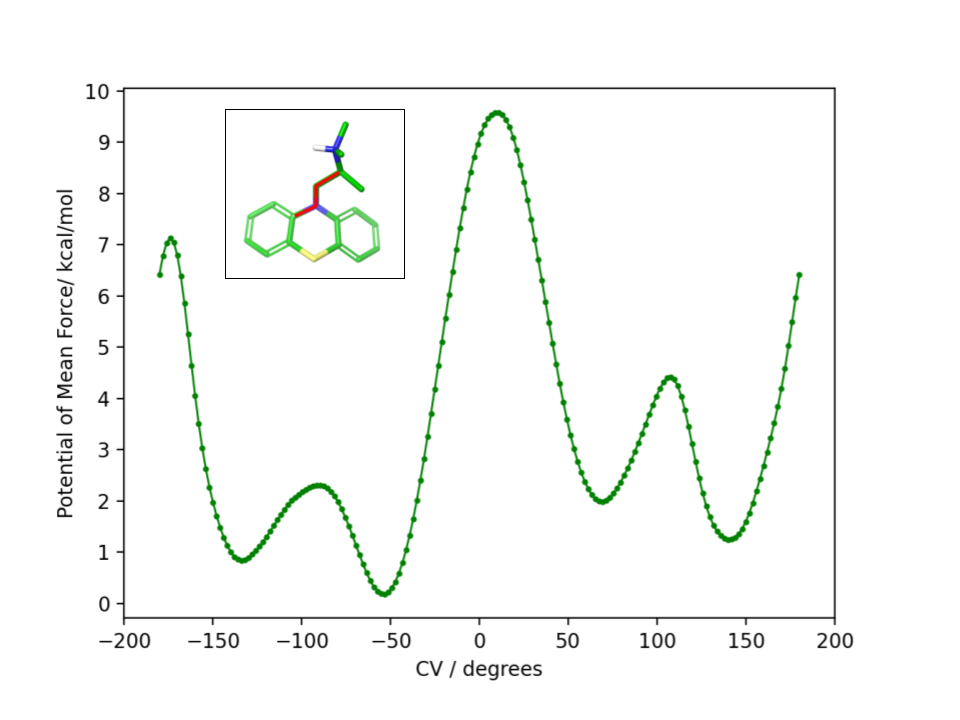}
    \caption{The potential of mean force (PMF) in water solution along the highlighted torsional angle, $\varphi$, of PMT1 computed by well-tempered metadynamics sampling.\cite{barducci2008well} The PMF identifies two major gauche conformational states at positive and negative angles around $60^\circ$ and $120^\circ$ separated by a large free energy barrier at $180^\circ$ of more than 7 kcal/mol from the global minimum at $-60^\circ$. The free energy barrier is sufficiently high that interconversions between the two stable conformational states are not observed in the time-scale of the alchemical calculations without the metadynamics landscape-flattening potential.\label{fig:pmf-torsion}} 
\end{figure}

Analysis of the molecular dynamics trajectories later revealed that the observed scatter of relative binding free energy estimates was due to the conformational trapping of the PMT guest in the starting conformation, which was randomly set during the system setup. Simulations with PMT trapped in some conformations overestimated the RBFE while those in the other underestimated it. We pin-pointed the conformational trapping to the branched alkylamine side chain of PMT which showed hindered rotation around one of its torsional angles (Figure \ref{fig:pmf-torsion}) caused by a large free energy barrier separating the gauche(+) and gauche(-) conformers (Figure \ref{fig:pmf-torsion}). The variations of conformers in the alchemical calculation broke the symmetry between equivalent poses and caused the scatter in the observed RBFEs. 

 To correct these inconsistencies, we modified our alchemical binding protocol to include a metadynamics-derived flattening potential bias that reduced the magnitude of the free energy barrier separating the conformers of the alkylamine sidechain of PMT (see Methods and Computational Details). We confirmed that the biasing potential was able to successfully induce rapid conformational transitions between these conformers, which were rarely achieved with the conventional ATM protocol. Consequently, integrating metadynamics-enhanced sampling with ATM (ATM-MetaD) indeed produced much better convergence of binding free energy estimates of symmetric poses starting from different initial conformers (Table \ref{tab:dgpmt1}). For example, the large discrepancy of RBFE estimates between the `spS` and `spR' binding poses was reduced to less than 1 kcal/mol with ATM-metaD and in closer consistency with statistical uncertainties. With only one exception, similarly improved convergence was achieved for the equivalent binding poses of bCD and mCD falling within a 1 kcal/mol range of each other (Table \ref{tab:dgpmt1}).

\begin{table}
\caption{\label{tab:dgpmt1} Relative binding free energy estimates of the binding poses of PMT relative to the same binding pose of PMZ for the two cyclodextrin hosts bCD and mCD and with and without metadynamics enhanced sampling.}
\begin{center}
\sisetup{separate-uncertainty}
\begin{tabular}{l S[table-format = 3.2(2)] S[table-format = 3.2(2)]}
 Pose & \multicolumn{1}{c}{$\Delta\Delta G_b$(ATM)$^{a,b,c}$} & \multicolumn{1}{c}{$\Delta\Delta G_b$(ATM+MetaD)$^{a,b,d}$} \\ [0.5ex] 
\hline
\multicolumn{3}{c}{bCD} \\
spS & 3.94(39) & 0.44(25) \\ 
ssS & 2.80(39) & 0.58(25)  \\ 
spR & 0.28(34) & 1.12(24)  \\ 
ssR & 4.62(45) & 1.65(25)  \\ \hline
psS & 1.98(36) & 1.49(24)  \\ 
ppS & 2.03(39) & 1.10(24) \\ 
psR & 0.77(29) & 1.46(24)  \\ 
ppR & 0.28(39) & 0.57(24) \\
\hline\hline
\multicolumn{3}{c}{mCD} \\
 spS & 1.93(37) & 0.96(25) \\ 
ssS & 0.57(44) & 0.11(26)  \\ 
spR & 1.59(41) & 0.30(25)  \\ 
ssR & 0.25(42) & 1.90(25)  \\ \hline
psS & 1.26(39) & -0.14(24)  \\ 
ppS & 0.20(42) & 0.95(25) \\ 
psR & 1.54(41) & -0.41(25)  \\ 
ppR & 0.10(38) & 0.10(24)  \\ 
\end{tabular}
\begin{flushleft}\small
$^a$In kcal/mol. $^b$Errors are reported as twice the standard deviation. $^c$Estimates  computational predictions submitted to the SAMPL9 challenge organizers. $^d$Revised ATM estimates with metadynamics conformational sampling. 
\end{flushleft}
\end{center}
\end{table}

\section{Discussion }

Molecular binding equilibria are central to applications ranging from pharmaceutical drug design to chemical engineering. Obtaining reliable estimates of binding affinities by molecular modeling is one of the holy grails of computational science. Enabled by recent developments in free energy theories and models, and an increase in computing power, early static structure-based virtual screening tools, such as molecular docking, are increasingly complemented by more rigorous dynamical free energy models of molecular recognition that represent the conformational diversity of molecules at atomic resolution. However, many challenges still remain to achieve a sufficient level of usability and performance for free energy models to widely apply them to chemical research. By offering a platform to assess and validate computational models against high-quality experimental datasets in an unbiased fashion, the SAMPL series of blinded challenges have significantly contributed to the advancement of free energy models.\cite{mobley2017predicting} By participating in SAMPL challenges we have refined and improved our models against host-guest and protein-ligand datasets\cite{Gallicchio2012a,Gallicchio2014octacid,GallicchioSAMPL4,deng2016large,pal2016SAMPL5,azimi2022application} and built an appreciation for the complexities of molecular recognition phenomena and the level of detail required to model them accurately. 

The present SAMPL9 bCD challenge highlights the importance of the proper treatment of conformational heterogeneity to obtain reliable quantitative descriptions of binding equilibria. We undertook this challenge with the mindset that host-guest systems are simpler surrogates of more challenging and conformationally diverse protein-ligand complexes and, hence, more suitable to assess computational methodologies. However, while working on setting up the molecular simulations, we realized that the phenothiazine/cyclodextrin complexes were far more chemically and conformationally diverse than expected. The majority of the guests exist in solution as mixtures of enantiomers related by nitrogen inversion (Figure \ref{fig:pheno-chirality}) which are distinctly recognized by the chiral hosts. As a result, one enantiomer could be significantly enriched relative to the other when bound to the host. In addition, each enantiomer binds the host in four generally distinct binding orientations that differ in the placement of the alkylamine sidechain and the aromatic substituent relative to the host (Figure \ref{fig:poses}). While in the experimental setting the guests and the complexes rapidly transition from one pose to another, this level of conformational heterogeneity poses serious challenges for standard molecular dynamics conformational sampling algorithms which are generally limited to one binding pose.

When facing these complexities, it is tempting to limit the modeling to the most important binding pose.  While it is true that often the most favorable pose contributes the most to the binding affinity and that neglecting minor poses results in small errors, binding pose selection remains an unresolved issue. Clearly, the identification of the major pose cannot be carried out by binding free energy analysis of each pose because that is precisely what one seeks to avoid in such a scenario. Whichever approach is adopted, it must be capable of identifying the most stable pose of each complex among many competing poses. The present results illustrate this challenge. For example, the populations derived from our free energy analysis (\ref{fig:bmode-populations}) indicate that the `spR' binding pose is often one of the most populated (red in Figure \ref{fig:bmode-populations}). However, CPZ is predicted to strongly prefer the `spS' pose when binding to mCD (orange in Figure \ref{fig:bmode-populations}B), and limiting the modeling to the `spR' pose would result in a gross underestimation of the binding free energy. Similarly, the TDZ-bCD complex is predicted to exist in a variety of poses (Figure \ref{fig:bmode-populations}A), including, for instance, the `psR' pose with the alkylamine sidechain pointing towards the primary face of bCD, with the `spR' pose contributing only a small fraction of the observed binding affinity. Clearly, at least in this case, limiting the modeling to one carefully selected predetermined pose would lead to significant mispredictions for individual complexes.

To obtain an estimate of the observed binding constants, in this work, we opted to compute the binding free energies of all of the relevant binding poses of the system and to integrate them using the free energy combination formula [Eq.\ (\ref{eq:combination-formula})]. The combination formula requires the populations of the conformational states of the guest in solution that, in this case, are easily obtained by symmetry arguments. Still, the work involved 64 individual alchemical free energy calculations (Figure \ref{fig:alchemical-process}) and hundreds of nanoseconds of simulation on GPU devices. While we attempted to automate the process as much as possible, setup errors were made and it is likely that some yet undiscovered defects are still affecting our revised results. We assessed the convergence of the pose-specific binding free energy estimates by monitoring the consistency between the results for symmetric poses. As a result of this assessment, we realized that one guest (PMT) was affected by slow conformational reorganization that required metadynamics treatment. This best-effort attempt resulted in good quantitative predictions for the complexes with $\beta$-cyclodextrin but, for largely unknown reasons, failed to properly describe the binding free energies of the complexes with its methylated derivative.

\section{Conclusions}

In this work, we describe our effort to obtain alchemical computational estimates of the binding constants of a set of phenothiazine guests to cyclodextrin hosts as part of the SAMPL9 bCD challenge using the Alchemical Transfer Method. The free energy modeling of these systems proved significantly more challenging than expected due to the multiple conformational states of the guests and the multiple binding poses of the complexes which had to be treated individually. Overall, 64 individual alchemical calculations were employed to obtain binding free energy estimates comparable to the experimental observations. The predictions were quantitative for the $\beta$-cyclodextrin host but failed to accurately describe the observed binding affinities to the methylated derivative. The work shows that even simple molecular systems can require extensive modeling efforts to treat conformational heterogeneity appropriately and it illustrates the role that multiple binding poses can play in protein-ligand binding prediction and, ultimately, drug design.

\section{Acknowledgements}

We acknowledge support from the National Science Foundation (NSF CAREER 1750511 to E.G.). Molecular simulations were conducted on Roivant's computational cluster and on the Expanse GPU supercomputer at the San Diego Supercomputing Center supported by NSF XSEDE award TG-MCB150001. We are grateful to Mike Gilson for providing experimental data for the SAMPL9 bCD challenge. We appreciate the National Institutes of Health for its support of the SAMPL project via R01GM124270 to David L. Mobley.

\section{Data Availability}

Input files of the AToM-OpenMM simulations are available on the GitHub repository {\tt github.com/\-GallicchioLab/\-SAMPL9-bCD}. The AToM-OpenMM software is freely available for download on GitHub.\cite{AToM-OpenMM} A detailed list of the results and their analysis are provided in the Supplementary Information. Simulation MD trajectories are available from the corresponding author upon reasonable request.

\section{Supplementary information}

Spreadsheets \href{https://docs.google.com/spreadsheets/d/1ftXvRjq36rh_LcUNihMgoCRSC1s_9XK-XTu3ZCDVmPE}{SAMPL9-bCDpc-FFEngine} and \href{https://docs.google.com/spreadsheets/d/1GOFku4EadRT9fpTbrfSbj7HYDC-6kx8Smq5JaLD6c8s}{SAMPL9-mCDpc-FFEngine} containing: (i) the absolute binding free energy of host G1 in the `s' and `p' binding modes, (ii) the relative binding free energies between G1 in the `p' and `s' poses and MTZ in the `ss', `sp', etc. binding modes, (iii) the relative binding free energy between MTZ and PMZ in each of the eight binding modes, (iv) the relative binding free energies between PMZ and the other guests in each of the eight binding poses, (v) the binding mode specific binding constants for each complex in each binding mode, and (vi) the calculated populations of each binding mode for each complex.

\appendix

\section{Proof of Eq.\ (\ref{eq:p1})}

Consider the potential energy function $U_0(x)$ of the uncoupled state of the receptor-ligand complex and $U_1(x)$ the one corresponding to the coupled state. Here $x$ represents, collectively, the degrees of freedom of the system. The probability that the complex to in binding mode  $i$ is
\begin{equation}
P_1(i) = \frac{\int_i e^{-\beta U_1(x)}}{\int e^{-\beta U_1(x)}}
\label{eq:p1-a}
\end{equation}
where the denominator is the configurational partition function of the complex in the coupled state, and the numerator, where the integration is restricted to regions of conformational space corresponding to binding mode $i$, is the configurational partition function of binding mode $i$ in the coupled state. Next, multiply and divide Eq.\ (\ref{eq:p1-a}) by the partition function $\int_i dx \exp[-\beta U_0(x)]$ of binding mode $i$ in the uncoupled state, noting that:
\begin{equation}
\frac{\int_i e^{-\beta U_1(x)}}{\int_i e^{-\beta U_0(x)}} = K_b(i)
\label{eq:p1-b}
\end{equation}
where $K_b(i)$ is the binding mode-specific binding constant.

To obtain an expression for the reminder ratio of the partition function of binding mode $i$ in the uncoupled ensemble to the partition function of the complex in the coupled ensemble, multiply and divide by the partition function of the system in the uncoupled ensemble $\int dx \exp[-\beta U_0(x)]$ noting that:
\begin{equation}
\frac{\int_i e^{-\beta U_0(x)}}{\int e^{-\beta U_0(x)}} = P_0(i)
\label{eq:p1-b2}
\end{equation}
where $P_0(i)$ is the population of binding mode $i$ in the uncoupled ensemble and
\begin{equation}
\frac{\int e^{-\beta U_1(x)}}{\int e^{-\beta U_0(x)}} = K_b
\label{eq:p1-c}
\end{equation}
is the overall binding constant. Collecting the terms above yields (\ref{eq:p1}).

\bibliographystyle{unsrt}
\bibliography{main, add2main}

\end{document}